\documentclass{elsart}
\def\bge{\begin{equation}}
\def\ede{\end{equation}}

\begin{document}

\begin{frontmatter}

\title{A CRITERION FOR THE CHOICE OF THE INTERPOLATION
KERNEL IN SMOOTHED PARTICLE HYDRODYNAMICS}

\author{Roberto Capuzzo--Dolcetta}
\address{Istituto Astronomico, Universit\`a di Roma ``La Sapienza", via G.M. Lancisi 29,I-00161, Roma, Italy.
e-mail: dolcetta@axrma.uniroma1.it
}
\author{Roberto Di Lisio}
\address{Dipartimento di Matematica, Universit\`a di Roma
``La Sapienza", P.le A. Moro 5, I-00100, Roma, Italy.
e-mail: dilisio@axcasp.caspur.it}
\end{frontmatter}
\vfill Keywords: interpolation procedures, particle methods, smooth particle hydrodynamics: kernels.
\vfill\eject

 {\bf Abstract.}  We study the problem of the appropriate choice
of the interpolating kernel to be used in the evaluation of gradients of
functions. Such interpolation technique is often used in applications, e.g. it is
typical fo  Smoothed Particle Hydrodynamics (SPH). We propose a minimization procedure for selecting
kernels in $\Re^n$, in the class of regular, normalizable, symmetric 
functions having finite moments up to a sufficiently high order; the method is valid  when
the kernel width is position--dependent and allows to recover conservation laws at the same order of approximation that SPH, as interpolation technique, has when the kernel size is constant.
{}{}

\vspace*{-0.5pt}
\noindent

\section{Introduction}

The aim of this paper is giving a criterion for the choice of
kernel functions to be  used in a particular interpolation problem.
The problem is the fitting of spatial derivatives of a function $f$
(for example the gradient of $f$, $\nabla f$)
known on a finite set of points in $\Re^3$.
The results here presented 
 do not depend on the way the points are chosen and valid in fully n--dimensional situations.

The problem of gradient interpolation 
assumes the form of the right choice of interpolating kernels in
the so-called Smoothed Particle Hydrodynamics (SPH) method
(we refer to [1] for
an introduction to this numerical method and to [2] for some recent
improvements).
In this scheme the fluid
is sampled by a set of
massive particles which are ``fluid representatives":
the particle positions  are considered
as moving
grid-points for the interpolation of the physical quantities characterizing
the fluid as well as
their spatial derivatives. SPH is so a fully Lagrangian method.
Analogously to the classic mollification approximation of a
sufficiently regular
function $f$ with the convolution function
$\langle f \rangle \equiv f*\phi$, where
$*$ is the convolution operator and $\phi$ is a kernel function
(whose properties will be specified later),
the spatial derivatives of $f$, resumed by
$\nabla f$, are approximated by $\langle \nabla f \rangle_1
\equiv (\nabla f ) * \phi$. In doing this, we take the advantage that,
under suitable
hypotheses on $f$ and $\phi$, $(\nabla f)*\phi$ can be replaced by
$f*\nabla \phi$. As final step, 
the integral $f*\nabla \phi$ is numerically evaluated.
Note that the knowledge of $f$
on a set of points suffices to evaluate $\langle \nabla f \rangle_1$,
and the
numerical evaluation of the integral giving $f*\nabla \phi$ can be
done using a Monte-Carlo estimate.
The interpolation procedure used in SPH follows the scheme we have just
sketched, so we shall refer to this scheme as ``SPH-like".
We start our analysis noting that another approximation for $\nabla f$
can be
given by $\langle \nabla f \rangle_2 \equiv \nabla(f*\phi)$.
As the kernel $\phi$ does not depend explicitely on the position, the
two approximations $\langle \nabla f \rangle_1$ and
$\langle \nabla f \rangle_2$ are equivalent
for any regular enough kernel.
It is easy to verify
that this property is lost when the kernel is position--dependent.

In the applications, to improve spatial resolution the characteristic size of the kernel is actually allowed to depend 
on the position and the
approximation $\langle \nabla f \rangle_1$ is implicitely applied. In this
case the SPH scheme is no longer 
momentum and energy conservative,[3].
In fact, the use of a position dependent kernel introduces, in the
approximation $\langle \nabla f \rangle_1$, a source of error proportional
to the characteristic size of the kernel while the approximation of
the SPH scheme is of the second order with respect to it, [1].

 To overcome this problem, the discrete SPH equations should be modified
introducing corrective terms as, for example, deeply  discussed 
in [4]. In [4] it is shown how the inclusion of the proper $\nabla h$ terms in the equations of motion leads to significant improvement of energy conservation in cases of SPH simulations of
collisions of polytropic spheres. 
\par\noindent In this paper we approach this problem in another way trying to answer to the question:

$\bullet$ is it possible to reduce  unconservativity within the general order of approximation of SPH without modifying its, appealingly simple, form?

We believe that the use of the approximation $\langle \nabla f \rangle_2$
would be desirable for this, and 
we will give, in the Appendix, a strong argument in support. 
Consequently, we simply look for kernels such that
the exchange between the approximation $\langle \nabla f \rangle_2$
and $\langle \nabla f \rangle_1$ introduces a negligible
error. This approach has the advantage that no changes on the discrete
SPH equations are needed.

We stress that the kernels proposed and commonly adopted
in SPH, have
been selected either following the spline interpolation theory,
[5,6], or
minimizing the error in the approximation of $f$ with $f*\phi$,
[7]. A recent paper, [9], gives a quite 
complete statistical analysis of the quality  of kernels of various
types (i.e. shapes) to reproduce the derivatives of given functions
sampled on a one-dimensional, even spaced, mesh.
Then our analysis is substantially different from all the previous
ones. In particular we are not interested here in the topic of
selecting the kernel's width as a function of
the particle's position, but rather to identify kernels able to keep second order accuracy of SPH as interpolation technique even in the case of position--dependent kernel width.

The plan of the paper is the following: in Section 2 we introduce
the kind of kernels we deal with; in Section 3 the
interpolation
procedure is studied in detail, and we state a general
criterion for selecting kernels; in Section 4 some 
kernels
are accordingly proposed; 
in Section 5, a simple numerical test is performed, while Sect. 6 is devoted to draw the Conclusions.

\section{The kernel function}
\noindent

First of all let us define the kernel $\phi$ to use in the
mollification procedure. We shall assume
the following general form
\bge
\phi({\bf r},h) \propto h^{-n} g(r/h)
\label{kernel}
\ede
where ${\bf r} = (r_1,r_2,...,r_n)$ is the position vector in $\Re^n$,
$r=|{\bf r}|$, $g(x)$ is a smooth
scalar function of the real variable $x$,
with $g'(0) \equiv (d
g/ dx)_{x=0}=0$ and such that the kernel has
finite moments up to a sufficiently high order.
The normalization condition $\int \phi~d{\bf r}=1$
(the integral is extended over all the physical space)
for any $h$ value is imposed.
Notice that these assumptions ensure that $\sup_{\bf r} | f*\phi -f| \propto h^2$, for any sufficiently smooth function $f(\bf r)$.
This means that the regularized approximating function $f*\phi$ tends
uniformly to the ``true" function $f$ as $h$ goes to zero.
The function $g$ defines the shape of the kernel, which is particularly
important when gradients have to be evaluated.

\section{A criterion for the kernel choice}
\noindent

In this Section we consider the step of the SPH-like interpolation
procedure related to the approximation of derivatives of a function.
In particular we study the difference
between
$(\nabla f)*\phi$ and $\nabla( f*\phi)$ when $\phi$ depends explicitly
on the position.
As a result we shall derive a first general criterion for the right
choice of
the shape
of kernels of the type (\ref{kernel}).

Before starting our analysis we find convenient to specify the
notations adopted hereafter. We define two differential operators,
\bge
\nabla_{\bf r} \equiv \left ({d\over dr_1}, {d \over dr_2}, ..., {d \over dr_n}
\right ),~~~~
\partial \equiv  \left ({\partial \over \partial r_1},
{\partial \over \partial r_2}, ..., {\partial \over \partial r_n} \right );
\label{diffoper}
\ede
the first one (also referred to as $\nabla$ when no ambiguity occurs) is the usual gradient operator in $\Re^n$ while the second one
is a ``partial" gradient operator,  in the sense that it
 deals 
only with the {\it explicit} dependence on the spatial variables.

As we have sketched in the Introduction, the basic idea behind the SPH-like
method is the approximation of the gradient
of a quantity basing on the knowledge of $f$
on the particle positions, only. This idea relies on that
$$
(\nabla f)*\phi = f*\nabla \phi= \nabla (f *\phi),
$$
these identities being true only under appropriate boundary conditions
for $f$ and $\phi$ and if $h=const$. When $h=h({\bf r})$ then integration by parts leads to
$\nabla( f * \phi) = f*\partial \phi$ and

$$
(\nabla f)*\phi \equiv \langle \nabla  f \rangle_1=\int \nabla_{\bf r'} f({\bf r'}) \phi({\bf r}- {\bf r'},h({\bf r})) d{\bf r'} =
$$

\bge =\nabla_{\bf r} \int f({\bf r'}) \phi({\bf r}- {\bf r'},h({\bf r})) d{\bf r'}
- \delta =
\nabla (f*\phi) -\delta \equiv \langle \nabla f \rangle_2-\delta,
\label{deltaerror}
\ede

where $\delta$ depends on $f$ and $h$.
In this case,
the normalization condition on the kernel
implies
\bge
\int \partial \phi({\bf r}-{ \bf r'},h({\bf r})) d{\bf r'}=0,
~~~~
\int
{d\over dh} \phi({ \bf r}-{\bf r'},h({\bf r})) d{\bf r'}=0.
\label{integrals}
\ede
Our aim is now to find how making $\delta$ as small as possible independently of $f$.
Taking into account (\ref{kernel}), let us evaluate
$\delta$:
$$
\delta \equiv \int f({\bf r'}) \nabla h({\bf r}){d\over dh} \phi({\bf r}-{\bf r'},h({\bf r}))
 d{\bf r'} \propto
$$
\bge
-{\nabla h({\bf r})\over h^{n+1}({\bf r})} \int f({\bf r'})
[n g({|{\bf r}-{\bf r'}|/h({\bf r})})
+ {|{\bf r}-{\bf r'}| \over h({\bf r})} g'(|{\bf r}-{\bf r'}|/h({\bf r}))]
d{\bf r'}.
\label{lowdelta}
\ede
In order to recover the equality
$(\nabla f)*\phi= \nabla (f*\phi)$, independently
of $f$, the
function $g$ in (\ref{lowdelta})
should satisfy, in $\Re^+$, the ordinary differential equation
\bge
n g(x) + x g'(x) =0.
\label{ode}
\ede
The non-trivial solution of this ODE is given by $g\propto x^{-n}$,
which leads to a non--normalizable kernel.
This means that, with a choice of $g$ in the set of normalizable kernels,
$\delta$ cannot vanish for all
functions $f$.

Thus, a possibility is to minimize $\delta$ via its expansion in terms of the parameter $h$, looking for conditions that
allow to eliminate at least some low order terms on $h$.

Performing a Taylor expansion of the function $f$ up to the second
order around the point $\bf r$, and
taking into
account the symmetry of the kernel and the normalizing conditions,
it is easy to verify that
\bge
\delta = {1\over 2}\nabla h({\bf r}) \Delta f({\bf r})
\int (r_1-{r'}_1)^2
{d \over d h}\phi({\bf r}-{\bf r'},h({\bf r})) d{\bf r'} + \psi({\bf r})
{\it O}(h^{2}),
\label{lowterms}
\ede
where $\psi$ is a function of the position and
$\Delta$ is the Laplace operator.
Now, we look for
kernels that minimize the absolute value of the integral in
(\ref{lowterms}). 
Let us work out the case $n\geq 3$ (the simpler cases
$n=1$ and $n=2$ must be handled in a slight different way).
Making the substitution ${\bf z}=({\bf r}-{\bf r'})/h({\bf r})$
the integral in (\ref{lowterms}) can be written as
\bge
h({\bf r}) \int z_1^2 ( n g(|{\bf z}|) + g'(|{\bf z}|) |{\bf z}|) d{\bf z}.
\label{cond1}
\ede
To perform the integral in (\ref{cond1})
we introduce the set of cylindrical variables
 $(z_1, \rho,\theta_3,...,\theta_n)$,
thus getting, after
simple manipulations of variables:
\bge
{1 \over 2} (n-1)\omega_{n-1}
\int_{-\infty}^{\infty} d\rho \int_{-\infty}^{\infty} |\rho|^{n-2}
z_1^2 \left [
 n g(\sqrt{z_1^2+\rho^2}) + g'(\sqrt{z_1^2+\rho^2})
\sqrt{z_1^2+\rho^2} \right ] dz_1,
\label{cond2}
\ede
where $\omega_n$ is the volume of the n--dimensional
unitary sphere.
Moreover,
using polar coordinates $(\alpha, \beta)$ on the plane
$(z_1,\rho)$, the expression (\ref{cond2}) becomes
\bge
{1 \over 2} (n-1)\omega_{n-1}\int_0^{2\pi}  sin^2(\beta) |cos(\beta)| d\beta  \int_0^{\infty}
\alpha^{n+1} (n g(\alpha) + g'(\alpha) \alpha) d\alpha;
\label{cond3}
\ede
the normalization condition, now, is
\bge
{1 \over 2} (n-1)\omega_{n-1}\int_0^{2\pi} |sin(\beta)| d\beta \int_0^{\infty} g(\alpha)
\alpha^{n-1} d\alpha=1.
\label{norm}
\ede

It is clear that the approximation of $(\nabla f)*\phi$
with $\nabla (f*\phi)$
improves of, at least, one order with respect to $h$
if the normalized kernel has a shape such that the
integral over $\alpha$ in
(\ref{cond3})
is zero. This improvement makes the SPH--like interpolation with variable kernel size of the same order (the second) of the classic SPH
interpolation with  constant $h$ (see Appendix).

Thus, we get a criterion to select kernels:

{\it
\noindent
kernels of type (\ref{kernel}) should be chosen selecting $g$
in the class of differentiable functions such that}
\bge
\int_0^{\infty} \alpha^{n+1} (n g(\alpha)+g'(\alpha)\alpha)~d\alpha=0~~,~~
0 < \int_0^{\infty} \alpha^{n-1} g(\alpha) d\alpha < \infty .
\label{criterion}
\ede
Remark 1: it is easily shown that also the cases $n=1$ and $n=2$ lead to the same constrain (12) for the selection of the kernel. Remark 2: the stated criterion is of course useless 
if $\nabla h({\bf r})$ is imposed to be zero.
Remark 3: this
analysis may be iterated by evaluating higher order terms in the
Taylor expansion of the interpolated function $f$, and imposing them to be zero.
\par As final, relevant, consideration we notice that the shape of
our proposed ~\lq optimal \rq~ kernel depends on the dimension, $n$, of the space ; of course this should be taken into account in the practical applications.

\section{Some examples and applications}
\noindent
We want, now, to apply the stated criterion to select kernels
(in some specific class of functions) in the most common three--dimensional case.  The
most used kernels in SPH numerical applications derive from the
theory of spline
functions, [5,6].
Thus, we consider the class of functions (which corresponds
to the third-order spline functions class) defined as:
\bge
g(r)=\left \{\matrix{ &ar^3+br^2+cr+d&~~~~~&0\leq r \leq 1&\cr
&a'r^3+b'r^2+c'r+d'&~~~~~&1\leq r\leq 2& \cr
& 0 &~~~~~& r > 2,&} \right .
\label{spline}
\ede
continuous up to the second derivative and such that:

\begin{enumerate}
\item $g(0)=1$ and $g(2)=0$;
\item the first derivative vanishes on the boundary of the
 set $[0,2]$.
\end{enumerate}

Following the criterion discussed in Section 3, the selected normalized $3-D$ kernel
in this class is found to be

\bge
\phi({\bf r},h)=
{15\over 1152 \pi h^3}
\left \{\matrix{ & ({r/ h})^2 \left (171~({r/h})-321 \right )+172&~~
&0\leq r/h \leq1
&\cr &&&& \cr
&\left (({r/ h})-2\right )^2\left (107-85~({r / h}) \right )
&~~&1\leq r/h \leq 2& \cr &&&& \cr
& 0 &~~& r/h >2. &} \right .
\label{selspline}
\ede

Another kernel often used in the applications belongs to the
super-Gaussian class of kernels:
\bge
g(r)=(ar^2+b)e^{-r^2}~~~~,~~~~ r \geq 0.
\label{sgauss}
\ede
When applied to this class of functions, our criterion yields (in $3-D$)

\bge
\phi({\bf r},h)= {1 \over {\pi^{3/2}h^3}}{\left ({5 \over 2}-(r/ h)^2 \right )
}e^{-(r/h)^2}.
\label{selgauss}
\ede

We note that
the kernel (\ref{selspline}) differs 
from the kernel belonging to the
same class and suggested by Monaghan and Lattanzio
[6] (see also Figure 1),
i.e.:
\bge
\phi({\bf r},h) = {1 \over \pi h^3 } \left \{
\matrix{& 1 - {3\over 2} (r/ h)^2 + {3 \over 4} (r/ h)^3
& ~~~ & 0 \leq r/h \leq 1& \cr
& & & & \cr
& {1\over 4} \left (2-(r /h) \right )^3 & ~~~
& 1 \leq r/h \leq 2 & \cr & & & & \cr & 0 & ~~~ & r/h > 2, &} \right .
\label{monlatt}
\ede
On the other hand
the kernel (\ref{selgauss}) is exactly the same super-Gaussian kernel
suggested
by Gingold and Monaghan [7].

\begin{figure}
\centerline{\vbox{\hrule width 5cm height0.001pt}}
\vspace*{1.4truein}     
\centerline{\vbox{\hrule width 5cm height0.001pt}}
\caption{The kernel (\ref{selspline}) selected with our method (solid line)
is compared with the kernel (\ref{monlatt})
(dashed line).
}
\end{figure}

Finally, note that a kernel which is somewhere negative
is compatible with our analysis, for we never made assumptions on the
sign of $g$ in (\ref{kernel}). Nevertheless, a non-negative kernel should be
desirable whether the interpolated function is thought to have some
specific physical meaning
(for example when
it represents a mass density). A possible way to overcome this
problem could be the
use of two different kernels to represent the physical quantities
and their gradients separately, but this point should deserve a deeper
analysis.

\section{A numerical test}
\noindent

\par 
 The problem of having a good interpolation of derivatives occurs  often
 in the applications and, even if it does not guarantee the quality of SPH evolutive simulations, surely helps in fluid--dynamical numerical simulations.
\par\noindent Of course, when violent shock fronts appear in fluid--dynamical
simulations, to keep energy and momentum conservations at a high level of precision requires both a large number of particles in around the front (i.e. high resolution) and modification of SPH equations by mean of the inclusion of the 
relevant $\nabla h$ term, as shown in [4]. Actually, we could check this in
an SPH simulation of the 
classic case of 1-D  shock tube calculation, after selecting the best  1-D  
kernel in the class
of cubic splines, following our, previously described, recipe. The treatment
 of shocks improves with our kernel, but the most significant improvement 
come actually from higher resolutions and $\nabla h$ terms inclusion.
\par
We shall use here  our selected kernel (\ref{selspline}) in a numerical example
of function gradient interpolation,
comparing our results with the analogous ones obtained using the kernel
(\ref{monlatt}) suggested by Monaghan and Lattanzio [6] (of course, the 
 3-D  case is by far the most important in practical applications).
To this aim, we evaluate numerically the gradient of a
function $f$, supposed known
on a set of positions, by SPH-like interpolation. We will perform the test
using kernels whose widths strongly depend on the position.  To this end
we will set the points on a non--uniform grid, and will fix the kernel
widths so that only a fixed number of points will be ``touched" by any kernel.

The integral $(\nabla f) * \phi \simeq \nabla (f*\phi)$
is then approximated, on the positions ${\bf r}_i$, by the Monte-Carlo
estimate
$$
{1 \over N} {\sum_{j=1}^N} \left (
{f({\bf r}_j) \partial_{\bf r} \phi({\bf r}
- {\bf r}_j,h({\bf r})) \over \rho({\bf r}_j)} \right )_{{\bf r}={\bf r}_i},
~~~~~~~i=1,\cdots,N
$$
where  $N$ is the number of positions ${\bf r}_j$ over which
$f$ is known and
$\rho$ is the number density of the points ${\bf r}_j$. 
As $f$ is assumed to be density of the points, the Monte-Carlo estimate
reduces to the simpler form
\bge
\nabla_{\bf r} f({\bf r}_i) \sim {1 \over N}{\sum_{j=1}^N}~ 
\left (\partial_{\bf r} \phi({\bf r} - {\bf r}_j,h({\bf r}))
\right )_{{\bf r}={\bf r}_i}.
\label{approxi}
\ede
In this case, the function $f$ itself is approximated as follows:
\bge
f({\bf r}_i) \sim {1 \over N}
{\sum_{j=1}^N}~ \phi({\bf r}_i - {\bf r}_j,h({\bf r}_i)).
\label{approxi2}
\ede
For the sake of comparison we perform the  numerical evaluations
using both the kernels (\ref{selspline}) and (\ref{monlatt}).

As ``test" function, we choose the function
$f(x,y,z)=(2|x-1/2|)^{-1/2}$
in the cube $0\leq x,y,z \leq 1$, assuming $f$ as
point density, too. Being $f$ diverging as $x \to 1/2$,
a lot of points ${\bf r}_j$ will fall in the neighbourhoud of $x=1/2$. 
The exact gradient is $\nabla f(x,y,z) = (
(x-1/2)/(2 \sqrt{2 |x-1/2|^5}),0,0)$, singular in $x=1/2$.
Taking into
account the meaning of $f$ as a density, 
the points are displaced on a regular grid. It is easy to verify
that the $i_x-th$
position is given by the root of the expression $1+\sqrt{2}\cdot (x-5)/\sqrt
{|x-0.5|}-2\cdot i_x/n_x$,
where $1 \leq i_x \leq n_x$ 
and $n_x$ is the number of
grid points
on the cube edge in the $x$ direction.
The setting on the other coordinates is trivial.
The total number of grid points
is so $n_x\cdot n_y\cdot n_z$.
In the test presented here $n_x=60$
and $n_y=n_z=50$. The fit to the function and to its gradient have been
obtained applying formulas (\ref{approxi2}) and (\ref{approxi}) respectively.

The kernel sizes $h$ are chosen such that for each point
the sums (\ref{approxi2}) and (\ref{approxi}) have only 2.5\% of
non-zero terms,
then the size of the kernel depends strongly
on the position. That is,
the width $h$ changes very rapidly in
space. This is surely a situation for which $\delta$, in equation (3),
differs from zero: this fact justifies the choice of this particular $f$. 

In Figure 2 the
fits obtained with the two cited kernels, along the
line $y=z=1/2$, are presented. Our kernel seems to
fits quite better either the function $f$ either the $x$ derivative.

\begin{figure}[htbp]
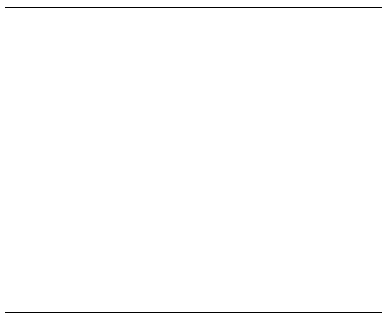

\vspace*{13pt}
\centerline{\vbox{\hrule width 5cm height0.001pt}}
\vspace*{1.4truein}     
\centerline{\vbox{\hrule width 5cm height0.001pt}}
\vspace*{13pt}
\caption{In Panel (a) the interpolation of the funcion $f$
(solid line) obtained using our kernel (dotted
line) and Monaghan and Lattazio's (dashed line) are shown;
Panel (b) refers to $\partial f / \partial x$.}
\end{figure}

As a final consideration, we note that, from a theoretical point af view,
the results of this test should depend mainly on the specific dependence
of the kernel's widths, $h=h({\bf r})$, with respect to the position, rather than on the particular function, $f$, chosen for the test.

\section{Conclusions}
\noindent
In this paper we have studied an interpolation problem
of great interest in practical applications. The problem is
the approximation
of the derivative of a function, known on a set of points, by a kernel
interpolating method. Given a suitable kernel function $\phi$,
the idea of the method consists in approximating
the derivatives
of a function $f$, and so its gradient,
with its mollified version $\langle \nabla f \rangle_1\equiv (\nabla f)
*\phi= f *\nabla \phi$,
and then evaluating numerically the last
convolution integral. This procedure fails as the kernel size depends
on the position, i.e. as the operation ``$*$" cannot be interpreted as a
convolution. In this case, we
stated a rule to choose the kernel.
The kernel is chosen such to control the difference between
$\langle \nabla f \rangle_1$ and the other possible mollification defined
as $\langle \nabla f \rangle_2 \equiv \nabla ( f * \phi)$ up to a given                      
order in $h$.
This simple criterion is independent
of the numerical method applied in the integration.
We have given some examples of 
kernels selected applying our criterion and discussed results  
in a typical case.

\bigskip
{\bf Appendix}

Let us consider the evolutive equations for a compressible fluid
written in conservative form, [8]:
\bge
\left \{
\matrix{&\rho_t + \nabla_{\bf r} \cdot {\bf q} =0 && \cr
& q_{it} + \nabla_{\bf r} \cdot \left ( {\bf q}{q_i \over \rho} \right )
+ \nabla_{r_i} P =0 &~~i=1,2,3;& \cr
& e_t+ \nabla_{\bf r} \cdot 
\left [ (e+P) {{\bf q} \over \rho} \right ]=0,&& \cr} \right .
\label{sisexact}
\ede
where ${\bf q}=\rho {\bf v}$ is the density of momentum, $P$ is the pressure,
$e/\rho= v^2/2+\varepsilon$
is the total energy per unit mass and $\varepsilon$ is the internal
one. 

The usual way to get the discrete SPH equations consists in the
convolution of each equation with a kernel $\phi$ and then in 
its approximation through
Monte-Carlo integrations, [1]. We shall consider here the
first step only. We begin considering a kernel of the type (\ref{kernel}) where
$h$ is a constant. The set of equations (\ref{sisexact}) becomes:
\bge
\left \{
\matrix{&\langle \rho \rangle_t + \nabla_{\bf r} \cdot
\langle {\bf q}  \rangle=0 && \cr
& \langle q_i \rangle_t + \nabla_{\bf r} \cdot \langle
\left ( {\bf q}{ q_i \over \rho} \right ) \rangle 
+\nabla_{r_i} \langle P \rangle=0 & ~~i=1,2,3;& \cr
& \langle e \rangle_t+ \nabla_{\bf r} \cdot \langle
\left [ (e+P) {{\bf q} \over \rho} \right ] \rangle=0,&& \cr} \right .
\ede
where $\langle \rho \rangle=\int \rho({\bf r'}) \phi({\bf r} - {\bf r'},
h) d{\bf r'}$, and so on. Let us study the conservation of the mollified
quantities. It is straightforward to verify that
${d \over dt} \int \langle \rho \rangle d{\bf r}=0$ and $\int \langle \rho \rangle d{\bf r}
= \int \rho$ d{\bf r}, and similar for the other quantities. 
The conservation of the mass derives directly by the equation
$(\nabla \cdot{\bf q}) *\phi = \nabla \cdot ({\bf q}*\phi)$ which means
(see the Introduction)
$\langle \nabla \cdot  {\bf q} \rangle_1 = \langle \nabla \cdot
{\bf q} \rangle_2$.
\par\noindent When
 $h=h({\bf r})$, then 
${d \over dt} \int \langle \rho \rangle d{\bf r}= O(h)$ and so on
(see Section
3); that is
the system is not longer conservative and the error derives from the
exchange of $(\nabla \cdot {\bf q}) * \phi$ with $\nabla \cdot ({\bf q}*\phi)
$. In other words, the error derives from the exchange of
$\langle \nabla \cdot {\bf q} \rangle_1$ with $\langle \nabla \cdot 
{\bf q} \rangle_2$.
The conclusion is that when the approximation $\langle \nabla f \rangle_2$
cannot be applied, the conservations are lost.

The use of kernels selected
on the basis of the criterion stated in Section 3 ensures 
${d \over dt} \int \langle \rho \rangle d{\bf r}={ O}(h^2)$, etc...,
which is the order of approximation  of the SPH method, [3].


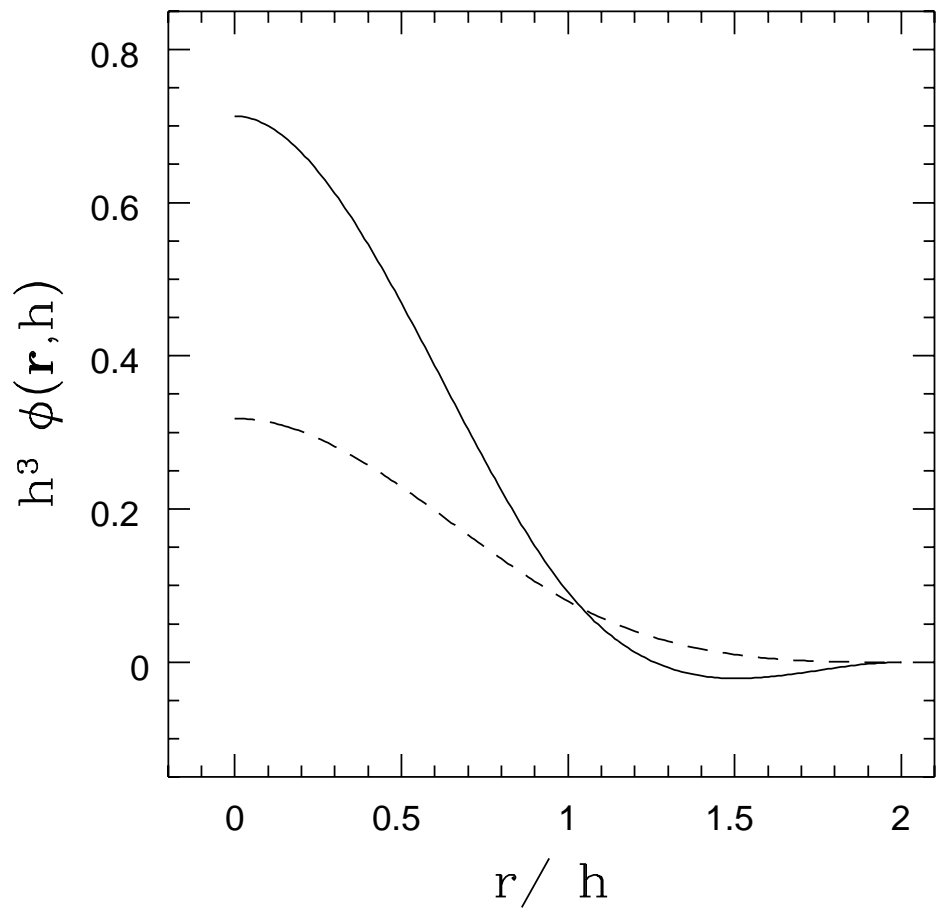
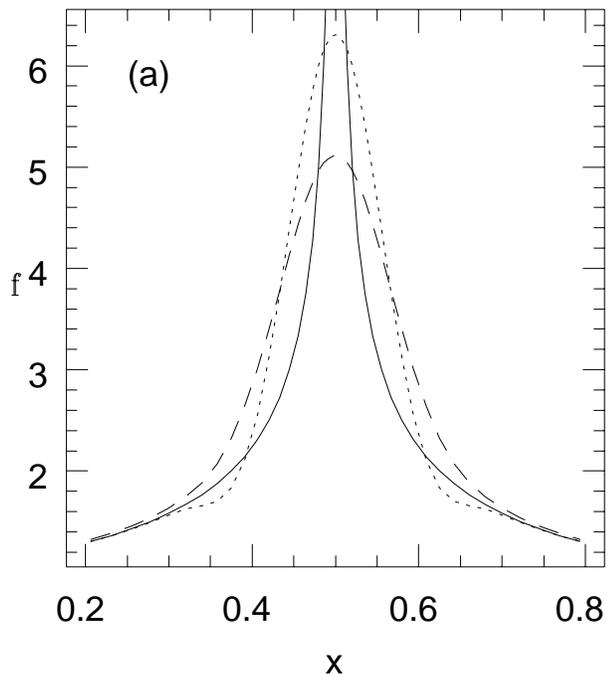
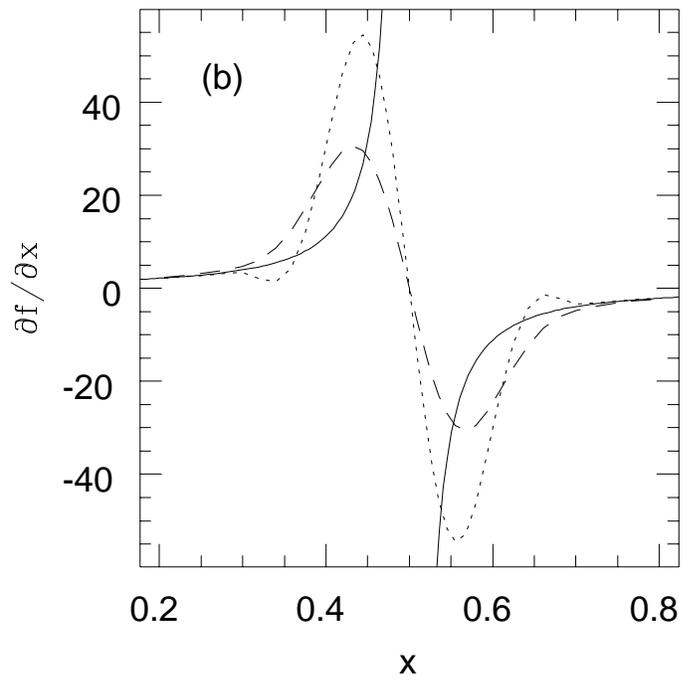
\end{document}